\documentclass[fleqn,usenatbib]{mnras}

\usepackage{newtxtext,newtxmath}

\usepackage[T1]{fontenc}

\DeclareRobustCommand{\VAN}[3]{#2}
\let\VANthebibliography\thebibliography
\def\thebibliography{\DeclareRobustCommand{\VAN}[3]{##3}\VANthebibliography}

\usepackage{graphicx}	
\usepackage{amsmath}	

\title{Imaging and spectroscopic observations of a confined solar filament eruption with two-stage evolution}

\author[Zhe Xu et al.]{
Zhe Xu$^{1,2}$\thanks{E-mail: xuzhe6249@ynao.ac.cn}
Xiaoli Yan,$^{1,2}$
Liheng Yang,$^{1,2}$
Zhike Xue$^{1,2}$
Jincheng Wang$^{1,2}$
Yian Zhou$^{1,2}$
\\
$^{1}$Yunnan Observatories, Chinese Academy of Sciences, Kunming 650216, China\\
$^{2}$Yunnan Key Laboratory of the Solar physics and Space Science,Kunming 650216, China\\
}

\date{Accepted XXX. Received YYY; in original form ZZZ}

\pubyear{2023}

\begin{document}
\label{firstpage}
\pagerange{\pageref{firstpage}--\pageref{lastpage}}
\maketitle

\begin{abstract}
Solar filament eruptions are often characterized by stepwise evolution due to the involvement of multiple mechanisms, such as magnetohydrodynamic instabilities and magnetic reconnection. In this article, we investigated a confined filament eruption with a distinct two-stage evolution by using the imaging and spectroscopic observations from the Interface Region Imaging Spectrograph (IRIS) and the Solar Dynamics Observatory (SDO). The eruption originated from a kinked filament thread that separated from an active region filament. In the first stage, the filament thread rose slowly and was obstructed due to flux pile-up in its front. This obstruction brought the filament thread into reconnection with a nearby loop-like structure, which enlarged the flux rope and changed its connectivity through the foot-point migration. The newly formed flux rope became more kink unstable and drove the rapid eruption in the second stage. It ascended into the upper atmosphere and initiated the reconnection with the overlying field. Finally, the flux rope was totally disintegrated, producing several solar jets along the overlying field. These observations demonstrate that the external reconnection between the flux rope and overlying field can destroy the flux rope, thus playing a crucial role in confining the solar eruptions.
 
\end{abstract}




\section{Introduction}

Solar filament eruptions are a complex phenomenon that often undergo a stepwise evolution \citep{kahler1992,stl2005,lr2007,yan2013,cx2020}. Previous studies have found that the filament eruptions produce a variety of solar activities as they develop into different stages, such as the flares, jets, and coronal mass ejections (CMEs) \citep{forb2000,zj2001,pf2002,cpf2011,hjc2011,stl2015,wyp2017,shen2021}. To better understand these solar activities, it is necessary for us to investigate the evolution of the filament eruptions, including their initiation, acceleration, and propagation processes, as well as their interactions with the surrounding structures at each phase. 

The kinematic evolution of filament eruptions usually consist of two stages: a slow-rise stage and a main-acceleration stage \citep{kahler1992,stl2005,shen2012,stl2016,cx2020}. In the first stage, the filament moves up slowly at a nearly constant speed of less than 50 km $s^{-1}$. In the second stage, the filament shows an exponential rise and its velocity was accelerated to greater than 100 km $s^{-1}$ in minutes \citep{stl2005,stl2011,cx2020}.The slow-rise stage often occurs before the main-acceleration stage and sometimes signals the onset of flares or CMEs \citep{zj2001,cx2023}. In most cases, there is a transition point between the slow-rise and main-acceleration stage, indicating that different driving mechanisms may operate in different stages and multiple physical processes may be coupled during the transition \citep{cx2020}. Therefore, how the filament eruption evolves from a slow-rise stage to a main-acceleration stage is still a topic of interest.

One possible mechanism for triggering and driving the filament eruption is the ideal MHD instabilities, such as the helical kink instability and the torus instability. The helical kink instability occurs when a twisted magnetic flux rope exceeds a critical value of twist \citep{fg2004,kli2004,tor2004}. The flux rope then becomes unstable and writhes, forming a distinct inverse-$\gamma$ or $\Omega$ morphology that can erupt \citep{jhs2003,ale2006,yjy2012,has2016}. On the other hand, the torus instability is related to the overlying magnetic fields that confine the twisted flux rope \citep{kli2006,dem2010,chan2017,cx2020}. If the overlying field decays sufficiently fast with height (called the decay index), the flux rope will undergo a rapid expansion. The torus unstable domain is usually determined by the critical height where the decay index reach a critical value \citep{kli2006}. 

Many filament eruptions with a stepwise evolution can be well explained within the framework of ideal MHD instabilities. \citet{fyh2005} found that the nonlinear kink motion of the flux rope reduces the confinement of the flux rope by changing its orientation at the apex.  \citet{lr2007} suggested that the catastrophic loss of equilibrium is induced by the kinking motions in the filament activation. \citet{vem2017} studied a prominence eruption with slow-rise and main-acceleration phases and attributed it to a kink unstable flux rope entering the torus unstable domain. In contrast, many failed filament eruptions are modeled as a kink unstable flux rope in the torus stable domain \citep{jhs2003,tor2005,has2016}.  \citet{cx2020} found that the main-acceleration stage starts at the critical height where torus unstable begins to dominate. \citet{chan2017} investigated a two-step filament eruption by calculating the decay index by the overlying field, and proposed that the evolution is due to the presence of successive instability-stability-instability zones as the filament rises. 

Previous studies have also revealed that magnetic reconnection plays a vital role in the filament eruption, such as the tether-cutting reconnection \citep{moore2001}, the emerging flux triggering mechanism \citep{cpf2000}, and the break-out reconnection \citep{ant1999}. Magnetic reconnection often occurs in multiple steps in a single event \citep{wyp2018,yjy2023,ylh2024}, which has been proposed as an alternative mechanism for the stepwise evolution of filament eruption \citep{kon2020,zqm2021,zrs2023}. Moreover, the magnetic field of the filament itself can also interact with the surrounding magnetic structures, which can either trigger or suppress the eruption of the filament. For instance, the interaction of a filament with nearby chromospheric fibrils and sheared loops can enlarge the filament or increase its twist, leading to its destabilization and subsequent eruption \citep{bi2012,chc2019,yan2020,yan2022,ylp2023}. Conversely, observations and simulations have demonstrated that external reconnection between an erupting flux rope and overlying field can reduce the upward hoop force of the flux rope, thus playing a vital role in confining the solar eruptions \citep{net2012,has2016,xue2016,kum2023,cj2023,jcw2023}. 

In this article, we investigate a confined filament eruption with a distinct two-stage evolution. This event was well observed by the Interface Region Imaging Spectrograph \citep[IRIS;][]{dep2014}, which operated the scanning slit moving roughly in path with the filament eruption. Combined with the multi-wavelength imaging observations from the Solar Dynamics Observatory \citep[SDO;]{pes2012}, this event provides an excellent opportunity for us to study the dynamic behavior of the filament in the stepwise eruption.

\section{Instruments and Data Reduction}
The data sets in this study are mainly from IRIS and SDO. IRIS obtains simultaneous spectra and slit-jaw images (SJIs) of the solar atmosphere with a spatial resolution of 0.$''$33-0.$''$4 and spectral resolution of 26 m\AA/53 m\AA\ for far-UV/near-UV passband over a field of view of 175$''$$\times$175$''$. The spectra of IRIS include three passbands, containing doublet bright lines of  Mg II h/k 2803/2796 \AA\ ($\sim 10^{4} K$, in the chromosphere), C II 1334/1335 \AA\ ($\sim 10^{4.3} K$, in the transition region), and Si IV 1394/1403 \AA\ ($\sim 10^{4.8} K$, in the transition region). IRIS/SJIs are obtained through four different wavelengths centered at 1330, 1440, 2796, and 2832 \AA. On 2021 December 24, IRIS observed the active region (AR) 12907 in the very large dense 320-step raster mode, which well covered the filament eruption here studied. IRIS/SJI 2796 \AA\  and 1400 \AA\ images are mainly used with a cadence of 37s. The spectra of the Mg II k line is used to diagnose the filament eruption with a step cadence of 9.3s. All the spectra and SJIs here used are obtained in the form of calibrated level 2, which is downloaded from the IRIS database.

SDO provides UV and EUV images from the Atmospheric Imaging Assembly \citep[AIA;][]{lem2012} and the magnetograms from the Helioseismic and Magnetic Imager \citep[HMI;][]{sch2012}. AIA images have a pixel size of 0.6$\arcsec$ and a cadence of 12/24s for the EUV/UV images. HMI data include the line-of-sight (LOS) magnetograms and the vector magnetograms with a pixel size of 0.5$\arcsec$. We mainly used the vector magnetograms with a cadence of 720s, which were remapped using a Lambert cylindrical equal-area projection and then transformed into the heliographic coordinates with the projection effect removed. All the data were differentially rotated to a reference time of 2021-12-24T10:13, close to the flare maximum.

\section{Results}

\subsection{Overview of the C4.2 flare SOL2021-12-24T10:13}

\begin{figure*}
	\centering
	\includegraphics[width=2.0\columnwidth]{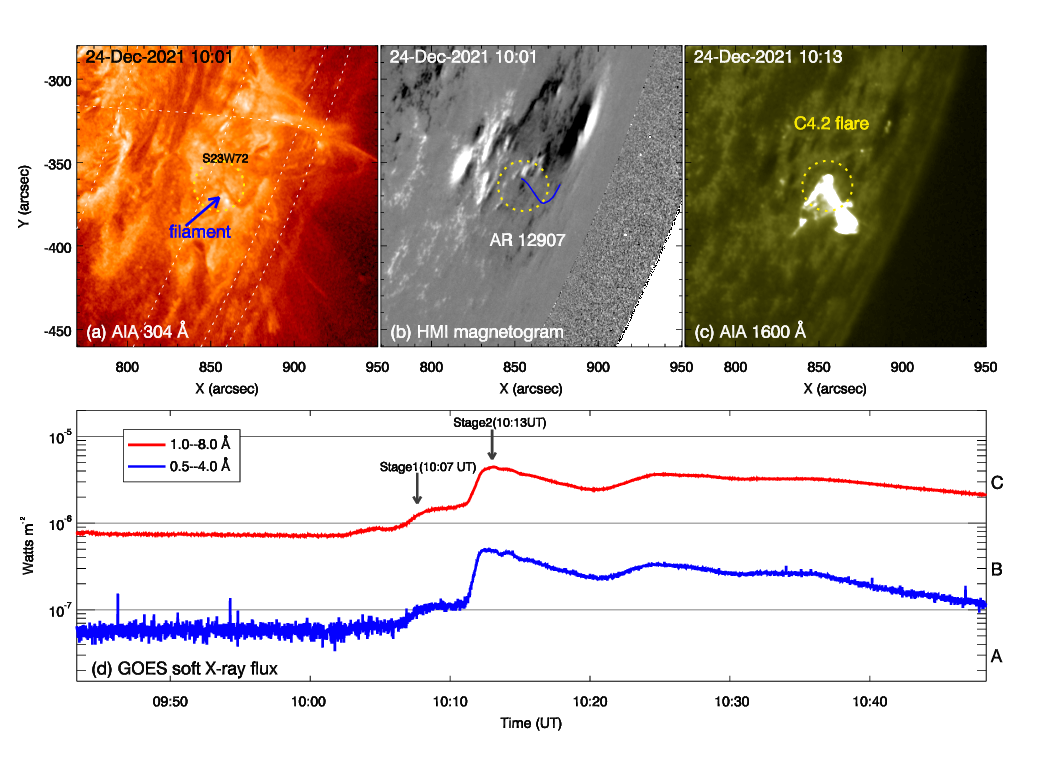}
    \caption{Overview of the solar flare SOL2021-12-24T10:13 (C4.2) in NOAA AR 12907. Panels (a)-(b): AIA 304 \AA\ image and HMI line-of-sight magnetogram showing the general configuration of AR 12907 before the flare, with the yellow dotted circle enclosing the main flare region. The blue arrow in Panel (a) points to the filament with impending eruption. Outline of the filament is also superposed as the blue curve in Panel (b). Panel (c): AIA 1600 \AA\ image showing the C4.2 flare at the peak time. Panel (d): Time profiles of the GOES soft X-ray 1-8 \AA\ and 0.5-4 \AA\ flux showing a two-stage evolution of the flare.}
    \label{fig1}
\end{figure*}

The solar flare in this article occurred in the NOAA active region (AR) 12907, located near the southwest limb of the solar disk (S23$^{\circ}$W72$^{\circ}$) on December 24, 2021. Figures \ref{fig1} (a) and (b) show the general view of the AR before the flare. A U-shaped filament can be seen in the middle of the active region, with its left leg lying above the neutral line of a small bipole. A few minutes later, the left part of the filament was activated and partially erupted, resulting in the flare captured by the AIA 1600 \AA\ images in Figure \ref{fig1} (c). An online animation \textit{onlinemovie1} is available to show the entire evolution of the filament eruption. According to the GOES 1-8 \AA\ soft X-rays flux (Figure \ref{fig1} (d)), this flare reached C4.2 class, and intriguingly, the GOES soft X-ray flux showed two apparent increases. The first one ended around 10:07 UT and the second one ended at 10:13 UT, which was the peak time of the flare. The soft X-ray flux curves suggest two intense energy release phases during the filament eruption, and it motivates us to investigate what mechanism caused the secondary acceleration of the eruption.

\subsection{First Stage: Initial Activation and Slow-rise Stage }

\begin{figure*}
	\centering
	\includegraphics[width=2.0\columnwidth]{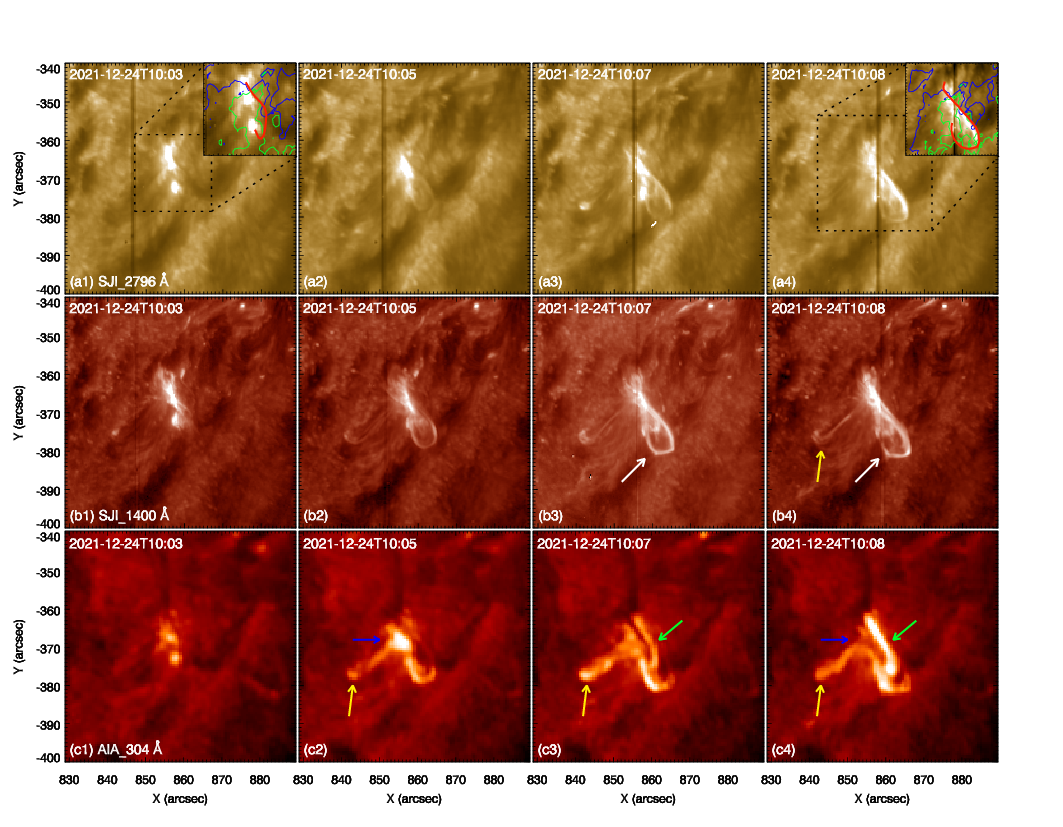}
    \caption{Evolution of the filament eruption in the first stage. Panels (a1)-(a4): Sequence of IRIS 2796 \AA\ images displaying the rising of an S-shaped filament in the chromosphere. In the close-up view at the top right corner of Panels (a1) and (a4), outline of the filament is highlighted by the red curve, and isogauss contours with levels of $\pm$200 G are overploted as blue and green contours. Panels (b1)-(b4): Sequence of IRIS 1400 \AA\ images showing the relevant brightening features in the transition region. Panels (c1)-(c4): Sequence of AIA 304 \AA\ images showing the interaction between the rising filament thread/initial flux rope (white arrows in (b3)--(b4)) and the ambient loop-like structure (yellow arrows).}
    \label{fig2}
\end{figure*}

Figure \ref{fig2} shows the sequence of IRIS 2796 \AA\ and 1400 \AA\ SJI images that display the chromospheric and transition region imaging of the filament eruption in the first stage.  At 10:03 UT, a thread-like structure suddenly activated and  separated from the left leg of the U-shaped filament. The separation motion was accompanied by the brightening of the two foot-points of the thread, as shown in Figures \ref{fig2} (a1) and (b1). In the highlighted partial view of Figure \ref{fig2} (a1), the filament thread was located around the magnetic neutral line, with its northern leg rooted in the positive polarity region and its southern leg rooted in the negative polarity region. The length of the filament thread was about 10 Mm, which was comparable to the scale of a mini-filament \citep{hjc2011,hjc2014,stl2015}. In the next 5 minutes from 10:03 to 10:08 UT, the filament thread rose and expanded southward, eventually evolving into an S-shaped flux rope structure, as shown by the partial highlighted view in Figure \ref{fig2} (a4). This S-shaped flux rope is sinistral since its axial magnetic field was left when viewed from the positive side of the PIL \citep{mar1998}. The transition of the filament thread from nearly straight to S-shaped indicated that the kink instability of the thread might be the main driving mechanism in this stage.

In the transition region, IRIS 1400 \AA\ images (Figures \ref{fig2} (b1)-(b4)) show that the ascent of the flux rope was accompanied by a significant brightening, indicating that the filament thread was heated during the slow-rise stage. The flux rope encountered an obstacle when it reached a certain height around (865'', -380''). The southern part of the S-shaped flux rope was squeezed and deformed, as indicated by the white arrow in Figures \ref{fig2} (b3) and (b4). This process can be seen more vividly in the online animation \textit{onlinemovie2}. Here, the obstruction was likely due to the magnetic flux pile-up in front of the rising flux rope, as evidenced by the deformation of the flux rope, which remained intact throughout the first stage. Furthermore, we observed that an ambient loop-like structure also showed a slight brightening following the ascent of the flux rope, as indicated by the yellow arrow in Figure \ref{fig2} (b4). In the AIA 304 \AA\ images (Figure \ref{fig2} (c1)-c(4)), the brightening of the loop-like structure was more apparent, and it first appeared around 10:05 UT. The most noticeable brightening occurred at the location where the loop-like structure came into contact with the rising flux rope, as marked by the blue arrow in Figure \ref{fig2} (c2). Subsequently, it can be seen that part of the loop-like structure merged with the flux rope at the contact point as shown in Figure \ref{fig2} (c4). Moreover, a twisted brightening structure wrapped around the filament thread was observed, as highlighted by the green arrow in Figure \ref{fig2} (c3) and (c4). This observation suggests that the exterior magnetic field wrapped around the filament thread underwent a magnetic reconnection with the surrounding loop-like structure during the slow-rise motion of the filament thread.

\subsection{Second Stage: Growth, Kinking, and Disintegration of the Filament} 

\begin{figure*}
	\includegraphics[width=2.0\columnwidth]{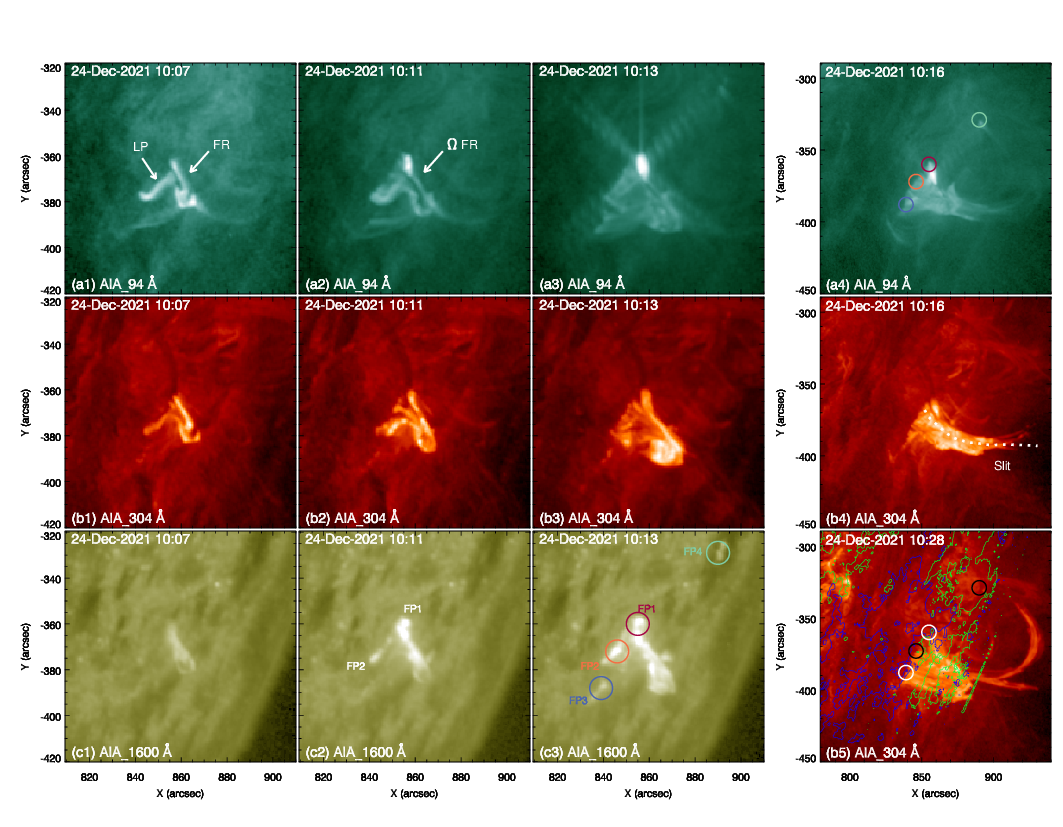}
    \caption{Evolution of the filament eruption in the second stage. AIA 94 \AA\ (Panels (a1)-(a4)), 304 \AA\ (Panels (b1)-(b5)), and 1600 \AA\ (Panels (c1)-(c3)) images showing the solar atmosphere at multiple temperatures. Panels (a4), (b4), and (b5) have a larger field-of-view to show more extended area involved in this eruption. The dotted curve in Panel (b4) denotes the slit position for the time-space digram in Figure \ref{fig5} (a). The four colored circles in Panel (c3) and (a4) mark the brightening location of the foot-point in 1600 \AA, and the white/black circle in Panel (b5) denotes the foot-point dominated by positive/negative polarity. Isogauss contours with levels of $\pm$200 G are overplotted as the blue and green contours in Panel (b5).}
    \label{fig3}
\end{figure*}

The transition of the filament eruption from the first stage to the second stage was well covered by the SDO/AIA multi-wavelength observations. Figure \ref{fig3} shows the sequence of AIA 94 \AA\ ($\sim$6.3 MK), 304 \AA\ ($\sim$~0.05 MK), and 1600 \AA\ ($\sim$~0.1MK) images that display the evolution of the solar atmosphere at multiple temperatures. At the end of the first stage (10:07-10:08 UT), the heated flux rope (FR) and the ambient loop-like (LP) structure were also observed in the AIA 94 \AA\ image, as pointed out by the white arrows in Figure \ref{fig3} (a1). The loop-like structure exhibited an S-shaped similar to the flux rope, suggesting that it was non-potential and might have a relatively strong twist. As mentioned before, FR and LP were under interaction at this moment. Then at 10:11 UT, FR and LP appeared to be fully connected and coalesced into a large $\Omega$-shaped flux rope, as seen in the AIA 94 and 304 \AA\ images (Figures \ref{fig3} (a2) and (b2)).  Meanwhile, significant brightening of two foot-points occurred in the AIA 1600 \AA\ image, as marked by FP1 and FP2 in Figure \ref{fig3} (c2). FP1 was in the region where the northern foot-point of FR was located, and FP2 was in the region where the eastern foot-point of LP was located. At this moment, FP1 and FP2 became the two foot-points of the $\Omega$-shaped flux rope, indicating that the interaction between FR and LP changed the connectivity of the erupting flux rope. This newly formed flux rope was unstable and continued to rise again, and then the filament eruption entered the second stage.

The newly formed flux rope demonstrated obvious rotational motion. As seen in Figures \ref{fig3} (a3) and (b3), the flux rope rotated clockwise about the vertical when viewed from above. The northwestern leg wrapped around to the back and the southeastern leg wrapped around to the front, transforming the $\Omega$-shaped flux rope into a cross-legged morphology. Such a morphology was usually considered a sign of a flux rope with increased kink instability \citep{ale2006,yjy2012,has2016}, referred to as the twist-to-writhe transition. The kink motion caused the flux rope to expand and rise further to a higher position. This allowed the erupting flux rope in direct contact with the overlying magnetic flux, as shown in Figure \ref{fig3} (a3). Meanwhile, remote brightening in FP3 and FP4 were also observed in the 1600 \AA\ image, as shown in Figure \ref{fig3} (c3). Based on the isogauss contours overplotted in Figure \ref{fig3} (b5), FP1 and FP2 were located near the magnetic neutral line connecting the erupting flux rope from positive to negative polarity, while FP3 and FP4 were rooted in the positive and negative polarity regions, respectively. It can be roughly distinguished from the AIA 94 \AA\ images that the overlying field mainly consisted of two bundles (Figures \ref{fig3} (a3) and (a4)). One bundle was on the southeast side of the flux rope and had FP3 as a footpoint. Another bundle was on the northwest side and had FP4 as a footpoint.  The footpoints on the opposite side of these two bundles of magnetic fields were obscured by the erupting flux rope. It is also likely that FP3 and FP4 were linked by a group of large-scale coronal loops as seen in the AIA 94 \AA\ images after the eruption (see the online animation \textit{onlinemovie1}).

After 10:13 UT,  the rising flux rope began to disintegrate and evolved into several solar jets along the overlying fields bundles, as seen in the AIA 304 \AA\ images (Figures \ref{fig3} (b4) and (b5)). The dominant jet moved along the southeast bundle toward a position around FP4, and a secondary smaller jet trailed behind it. Another jet along the southeast bundle toward a region near FP3 was also detected. Obviously, the filament eruption changed its character from a flux-loop-like to a jet-like at this stage. This transition is well consistent with the solar jet model by \citet{stl2015}, which suggests that the reconnection between the rising loop and the overlying field can produce a jet-eruption. Consequently, the rising flux rope was totally disintegrated after these jets and all the eruptions were confined by the overlying field bundles.

\subsection{Spectroscopic observations of the filament eruption}

\begin{figure*}
	\includegraphics[width=2.0\columnwidth]{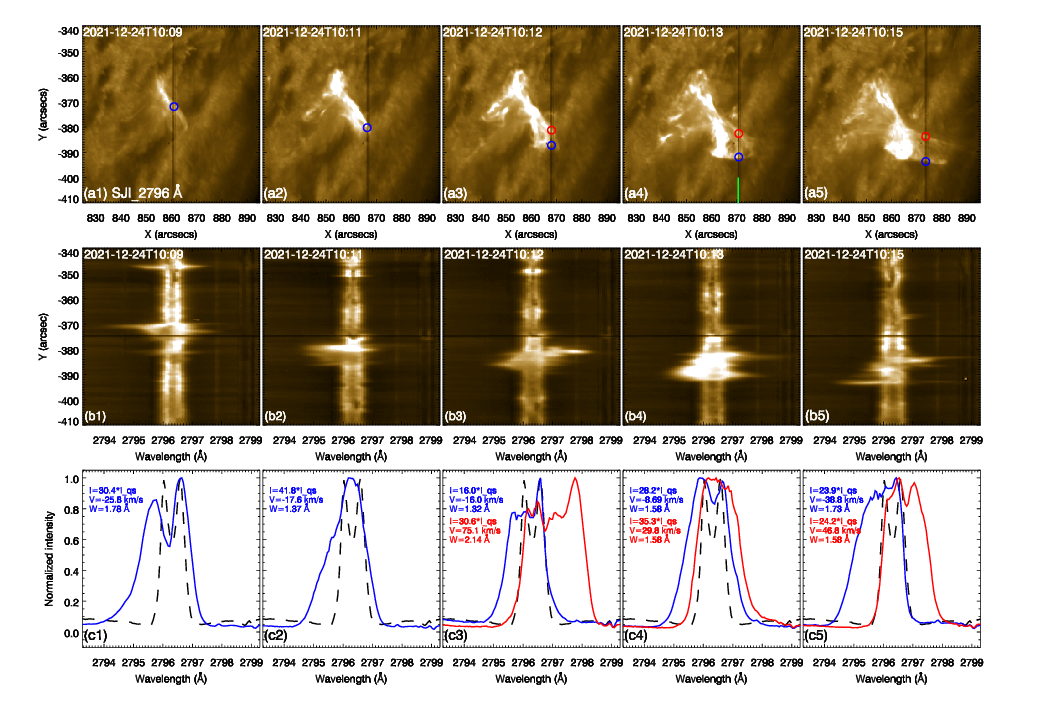}
    \caption{Spectroscopic observations on the filament eruption. Panels (a1)-(a5): Sequence of IRIS 2796 \AA\ SJI images displaying the twist-to-writhe transition of the erupting filament. Panels (b1)-(b5): Spectra of the Mg II k line at the corresponding slits shown in the top panels. Panels (c1)-(c5): Line profiles of the Mg II k line at the selected pixels where the scanning slit across the filament. The blue/red curve denotes the line profile of the pixel shown by the blue/red circles in Panels (a1)-(a5). The black dashed curve denoted the quiet spectrum of Mg II K line obtained by taking the median value from the quiet Sun indicated by the vertical green line in Panel (a4). }
    \label{fig4}
\end{figure*}

Figure \ref{fig4} and the online animation \textit{onlinemovie2} show the IRIS/SJI 2796 \AA\ imaging and spectra of the Mg II k line. The sequence of IRIS 2796 \AA\ SJI images (Figures \ref{fig4} (a1)-(a5)) clearly show the twist-to-writhe transition of the flux rope. The slit swept across the top of the flux rope where the rotational motion mainly occurred. Figures \ref{fig4} (b1)-(b5) show the spectra of the Mg II k line at the corresponding slits. Figures \ref{fig4} (c1)-(c5) show the spectral line profiles of selected pixels (marked by blue and red circles) and compares them with the quiet spectrum of the Mg II k line (dashed curves). The Doppler velocity of the Mg II k line in the selected pixel was measured using the center-of-gravity method with the quiet spectrum as the reference. The line width of the Mg II k line was computed by taking its full width at half maximum.

At 10:09 UT, the IRIS slit scanned the middle part of the flux rope. Owing to its chromospheric origin, the profile of the Mg II k line showed strong self-absorption in the line center due to the large opacity of the filament, as shown in Figures \ref{fig4} (b1) and (c1). A large Doppler blueshift of -25.8 km/s and a line width broadening of 1.78 \AA\ can also be observed. We speculated that the blueshift was caused by the ascent of the filament in the line-of-sight direction, and the line width broadening was caused by the magnetic reconnection below the filament in the first stage. At 10:11 UT, the slit scanned the top of the flux rope when the filament interacted with the ambient loop-like structure. The line intensity was strongly enhanced to 41.8 $I_{qs}$ (intensity at quiet Sun) and the self-absorption in the line center disappeared, as shown in Figures \ref{fig4} (b2) and (c2). This might indicate that the filament was heated to a temperature of transition region that reduced the opacity, as suggested by \citet{car2015}. The Doppler blueshift still existed at the top of the flux rope, but was slightly reduced to a velocity of -17.6 km/s.

In the next period, from 10:12 UT to 10:13 UT, the kinking motion of the flux rope was obvious when the slit swept across the top of the knotted flux rope. At this stage, the cross-legged morphology had formed. Its front leg began to rotate downward and the rear leg began to rotate upward along the line-of-sight. From the spectra of the Mg II k line (Figures \ref{fig4} (b3)-(b4)), one can see a significant redshift appeared in the north part (marked by the red circle), while it reversed to a blueshift in the south part ( marked by blue circle). The maximum of the red and blue shifts corresponded exactly to the north and south edges of the flux rope. Then at 10:15 UT, when the slit scanned the overlying flux bundle, the spectra of the Mg II k line also showed a clear red-blue asymmetry in Figure \ref{fig4} (b5). The line profiles of the selected pixels (marked the by red and blue circles) were extremely similar, with an intensity of $\sim$24 $I_{qs}$, Doppler velocities of 46.8 and -38.8 km/s, and line width broadenings of 1.58 and 1.73 \AA. This phenomenon suggests that the rotational motion was transferred from the flux rope to the overlying fields as the solar jet appeared.

\subsection{Time evolutions} 

\begin{figure*}
	\includegraphics[width=2.0\columnwidth]{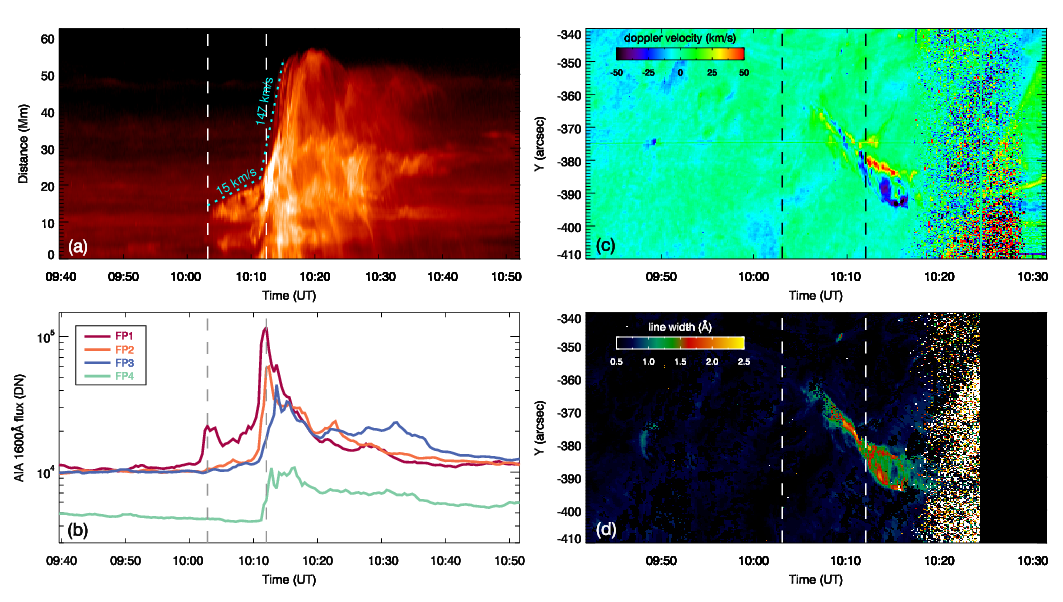}
    \caption{Time evolution of the solar filament eruption. Panel (a): Time-space diagram of AIA 304 \AA\ images along the slit as shown in Figure \ref{fig3} (b4), showing the filament eruption with a distinct two-stage evolution. The cyan dotted lines are the linear fitting to the front edge of the filament eruption, showing that the average rising speed of the filament in two stage is 15 km/s and 147 km/s, respectively. Panel (b): Light curves of the foot-points brightening in AIA 1600 \AA\ with colors denoting FP1-FP4 as shown in Figure 3 (c3).  Panels (c) and (d): Slit-time plots of Doppler velocity and line width of the Mg II k line during the filament eruption. Two dashed vertical lines in each panels denote the peaks of FP1 brightening in the first and second stages, respectively.}
    \label{fig5}
\end{figure*}

To study the temporal evolution of the filament eruption, a time-space diagram of the AIA 304 \AA\ images was made along the direction of filament eruption. The result is presented in Figure \ref{fig5} (a). Clearly, the kinematic evolution of the filament eruption can be recognized as two distinct stages. In the first stage, from 10:03 UT to 10:11 UT, the filament rose slowly with a projected average speed of 15 km $s^{-1}$. Then the filament accelerated in a very short time and the projected average speed reach a level of 147 km $s^{-1}$ in the second stage. The transition point between the first and the second stages was clear at 10:11 UT, which was exactly the time when the new $\Omega$-shaped flux rope completely formed, as shown in Figures \ref{fig3} (a2) and (b2).

The intensity evolutions of the foot-points (FP1-FP4) were further traced to demonstrate the temporal relationship between the filament eruption and the sequence of foot-point brightening. We found that the two brightening peaks of FP1 (shown by two vertical dashed lines at 10:03 UT and 10:12 UT) coincided with the onset times of the two stages of the eruption at 10:03 UT and 10:11 UT. Moreover, it was found that only FP1 showed significant brightening in the first stage, while the other FP2-FP4 had no obvious response. This suggested that the eruption in the first stage was induced by the small S-shaped filament thread and was confined in a compact region around FP1, as previously illustrated in Figure \ref{fig2}. At the beginning of the second stage, FP1 and FP2 started to brighten and peaked simultaneously at around 10:12 UT. It suggests that the eruption in the second stage was due to the Omega-shaped flux rope at a larger scale, which was formed by reconnection between the original flux rope and the ambient loop-like structure and rooted in FP1 and FP2, as shown in Figure \ref{fig3}. FP3 and FP4 also showed brightening in the second stage, but their onset and peak times were slightly delayed compared to FP1 and FP2. This delay demonstrate that the remote brightenings at FP3 and FP4 resulted from the magnetic reconnection between the new flux rope and the overlying field. The resulting destruction of flux rope and formation of solar jets along the overlying field also verify this reconnection procedure, as shown in Figure \ref{fig3}.

Doppler velocity and line width of the Mg II k line were also obtained along the scanning slit to show the dynamic behavior of the filament. Figures \ref{fig5} (c) and (d) show that the properties of the flux rope in the first and second stages were different. In the first stage, blueshift and line broadening were mainly concentrated in the middle of the flux rope. It suggests that the filament was dominated by ascending motion in the first stage. In the second stage, the flux rope showed an asymmetry of red-blue shift, manifesting as a redshift in the north part and a blueshift in the south part. This indicates that the rising flux rope was under rotational motion in the second stage, which conforms to the action of the kink instability. It was also found that the line width broadening occurred primarily at two time intervals, 10:09-10:11 UT and 10:13-10:16 UT. These two intervals corresponded to the two reconnection procedures, i.e., reconnection between the S-shaped flux rope and ambient loop, forming the $\Omega$-shaped flux rope, and the reconnection between the $\Omega$-shaped flux rope and overlying field, forming the jet.

\section{Conclusions and Discussions}

In this article, we presented a detailed analysis of a solar filament eruption by using the imaging and spectroscopic data from IRIS and SDO. The filament eruption showed a distinct two-stage evolution, which consisted of a slow-rise stage followed by a rapid eruption stage. The main driving mechanism of the filament eruption at each stage has been discussed. The observational results and our interpretations are concluded as following.

The filament eruption originated from a filament thread splitting from a U-shape filament. The filament thread rose slowly in the first stage and transformed from a nearly straight thread to an S-shaped flux rope. Such a procedure was very similar to the case reported by \citet{bi2015}. They investigated a partial filament eruption, which was induced by a filament thread splitting from the middle of the filament in a kinking fashion. Previous study has also shown that the kink instability was responsible for the slow-rise phase of the filament eruption \citep{vem2017}. In addition, no significant flux emergence or magnetic cancelation was observed in this stage, therefore the mechanism of magnetic reconnection can be ruled out. Accordingly, the kink instability of the filament thread was probably the main driving mechanism in the first stage.

The filament thread was obstructed and deformed when it rose to a certain height. The filament did not stop evolving, but began to reconnect with the ambient loop-like structure, resulting in the formation of a new, larger flux rope. This interaction led the erupting flux rope to change its connectivity, with one of its foot-points migrating from the southern side of the filament to the far side of the loop-like structure. The newly formed flux rope exhibited stronger kink instability, manifested by its top rotation to form a cross-legged shape. This process was similar to the filaments interaction previously reported by \citet{jyc2013}. They found that the failed eruption of a small filament brought it into contact with a nearby filament, and that the interaction between them created a new filament with a clear cross-legged shape. \citet{bi2012} also found that a filament thread driven by kink instability underwent a reconnection with the surrounding magnetic field, resulting in a more twisted field of the erupting filament. Moreover, growth of filaments caused by interaction with their adjacent structures were also reported \citep{yb2016,ylp2023}. Recently, \citet{gty2023} provided a case of an S-shaped thread developing into an erupting flux rope. They discovered that one of the sigmoidal thread's foot-points migrated to a completely different location in conjunction with a twofold rise in magnetic flux, and suggested that the magnetic reconnections between the flux rope and the surrounding magnetic fields led to the twist development of the flux rope. In our study, the growth of the flux rope, the migration of the foot-point, and the appearance of the cross-legged shape were all clearly observed. Therefore, we suggested that the interaction between the filament and the surrounding loop-like structures might enhance the kink instability of the erupting flux rope, which in turn triggered the second stage of the filament eruption. 

The kink instability drove the flux rope to rise into the higher atmosphere, which triggered the filament to interact with the overlying field. Afterwards, the filament eruption changed from a flux-loop-like to a jet-like morphology. This transition confirms to the solar jet model in \citet{stl2015}. They proposed that small-scale filament eruptions can be taken as the driver of solar jets, through the reconnection between the rising loop and the overlying field. In this event, several jet-eruptions emerged along the overlying field bundles, and all were accompanied by remote brightening. These phenomena imply that the external reconnection between the filament and overlying field took place in the second stage. As a consequence, the filament was completely disintegrated and the eruption was confined by the overlying field. In fact, external reconnection between the erupting flux and the overlying field has frequently been reported in the confined eruptions \citep{net2012,has2016,xue2016,kum2023,ylh2023}. \citet{cj2023} and \citet{jcw2023} modeled the external reconnection between the flux rope and the overlying magnetic fields and suggested that the reconnection can erode the flux rope, decreasing its upward hoop force, and eventually destroy the flux rope. In this event, the destruction of the flux rope and the resulting formation of jets provide an observational evidence to support this view.

Particularly, the availability of spectroscopic observation from the IRIS mission reveals several details about the dynamics of the filament in this study. The spectra of the Mg II k line changed from blue-shift dominance to red-blue asymmetry, indicating the rotation of the ascending flux rope, which conforms to the action of the kink instability. The reconnection between the flux rope and ambient loops in the second phase is evidenced by the line width broadening, intensity brightening, and the disappearance of self-absorption on the Mg II k line profile. The red-blue asymmetry of the Mg II k line of the overlying field also shows that the rotation was transferred from the flux rope to the overlying field as the solar jet emerged.

However, tracing the twist evolution of the flux rope during the eruption is challenging, since the magnet field measurements in this event were not very accurate due to the limb effect. Our discussions of the kink instability of the flux rope are limited to qualitative speculation based on the imaging and spectroscopic observations. In future, events with accurate magnetograms will help us confirm our idea that the interaction of flux rope with the surrounding structures can enhance the twist of flux rope, thereby increasing the MHD instabilities of the flux rope and ultimately contributing to the filament eruption. 

\section*{Acknowledgements}

The authors would like to express their gratitude to the referee for many valuable comments that helped improve this article. We are grateful to the IRIS and SDO science teams for providing the data. IRIS is a NASA small explorer mission developed and operated by LMSAL with mission operations executed at NASA Ames Research center and major contributions to down-link communications funded by the Norwegian Space Center (NSC, Norway) through an ESA PRODEX contract. This work is supported by the National Key R\&D Program of China (2019YFA0405000), the Strategic Priority Research Program of the Chinese Academy of Science, Grant No. XDB0560000, the National Science Foundation of China (NSFC) under grants 12203097, 12325303, 12173084, 12073072, 12273108, and 12273106, the Yunnan Key Laboratory of Solar Physics and Space Science under the number 202205AG070009, the Yunnan Science Foundation of China (202301AT070349 and 202301AT070347), the Youth Innovation Promotion Association, CAS (Nos 2019061 and 2023063), the CAS ``Light of West China'' Program, and the Young Elite Scientists Sponsorship Program by Yunnan Association for Science and Technology.

\section*{Data Availability}
The data used in this article are publicly available. The calibrated level 2 data of IRIS are downloaded from the IRIS database (\url{https://iris.lmsal.com/data.html}). SDO/AIA and SDO/HMI data are downloaded from the Joint Science Operation Center (\url{jsoc.stanford.edu}).


\bsp	
\label{lastpage}
\end{document}